# Relativistic Coupled-Cluster Study of Diatomic Metal-Alkali Molecules for Electron Electric Dipole Moment Searches

A. Sunaga[1]*, M. Abe[1], V. S. Prasannaa[2], T. Aoki[3], and M. Hada[1],
[1]*Tokyo Metropolitan University, 1-1, Minami-Osawa, Hachioji-city, Tokyo 192-0397, Japan*
[2]*Physical Research Laboratory, Atomic, Molecular and Optical Physics Division, Navrangpura, Ahmedabad-380009, India*
[3]*The University of Tokyo, 3-8-1 Komaba, Meguro-ku, Tokyo 153-8902, Japan*

Recent improvements in experimental techniques for preparing ultracold molecules that contain alkali atoms (e.g., Li, Na, and K) have been reported. Based on these advances in ultracold molecules, new searches for the electric dipole moment of the electron and the scalar-pseudoscalar interaction can be proposed on such systems. We calculate the effective electric fields ($E_{\mathrm{eff}}$) and the S-PS coefficients ($W_s$) of SrA and HgA (A = Li, Na, and K) molecules at the Dirac-Fock (DF) and the relativistic coupled cluster (RCC) levels. We elaborate on the following points: i) Basis set dependence of the molecular properties in HgA, ii) Analysis of $E_{\mathrm{eff}}$ and $W_s$ in SrA and HgA, and comparison with their fluoride and hydride counterparts, iii) Ratio of $W_s$ to $E_{\mathrm{eff}}$ ($W_s/E_{\mathrm{eff}}$) at the DF and the correlation RCC levels of theory.

## I. Introduction

The electric dipole moment of the electron (eEDM) is a physical property of the particle (if detected) that arises from Parity ($P$) and Time-reversal ($T$) symmetry violations [1,2]. Although the existence of the eEDM is predicted in the standard model (SM) of particle physics, its predicted value is extremely small (($|d_e| \approx 10^{-38}$ $e$-cm [3], $|d_e| \approx 10^{-40}$ $e$-cm [4]) and therefore, measuring its SM value is currently not possible. In contrast, many particle physics theories that are beyond the standard model (BSM) predict values of the eEDMs that are several orders of magnitude greater than their SM counterparts [2–5], and some are well within reach of current experiments [6–8]. Therefore, upper bounds on the eEDM placed by experiments, thereby constraining stringently several post-SM theories, are a crucial probe of BSM physics. In particular, eEDM tabletop experiments that use atoms and molecules can probe PeV energy scales, which are well beyond the reach of accelerators [3].

Another $P$, $T$ violating interaction, but which is predicted only by BSM theories, is the scalar-pseudoscalar (S-PS) interaction between the nuclei and the electrons in an atom or a molecule [9–11]. The coupling constant associated with this interaction is the S-PS constant ($k_s$). The S-PS interaction is predicted in, for example, the minimal supersymmetric standard model (MSSM) [12], where the loop-induced Higgs-gluon–gluon couplings contribute to $k_s$, and the aligned two-Higgs-doublet model (A2HDM) [13]. The S-PS interaction is not predicted in all BSM theories, but its importance relative to the eEDM depends on the theory. For example, there is a model which predicts a large contribution of the S-PS interaction to the atomic (and molecular) EDM as compared to the eEDM [14]. More details on the importance of eEDM searches and the S-PS interaction can be found in Chupp's review [15].

The values of $d_e$ and $k_s$ are obtained by a combination of experimental energy shifts in atoms or molecules and theoretically determined enhancement factors (further details can be found in Appendix A of Ref. [16]). The latter can be calculated only by using atomic or molecular relativistic many-body theories. The enhancement factor for the eEDM interaction is the effective electric field ($E_{\text{eff}}$), while that for the S-PS interaction is the S-PS coefficient ($W_s$). Since both the eEDM and the S-PS interactions contribute to the measured energy shift in an experiment, we need to perform measurements on two or more systems with different sensitivities to these interactions, in order to obtain their contributions individually (c.f. Figure 1 in Ref. [17] and Figure 1 in Ref. [18]). In the subsequent sections, we discuss the sensitivity of a given system in terms of the ratio between their $W_s$ and $E_{\text{eff}}$ ($W_s/E_{\text{eff}}$).

Molecules that can be cooled to the ultracold regime are attractive as candidate systems for an eEDM experiment because of their large coherence time and the total number of molecules that can be used for that experiment. One such set of molecules that offer promise for future eEDM search experiments are metal-alkali diatomic systems. In fact, several groups have successfully reported on the cooling of systems such as YbLi [19,20], HgRb [21], etc.

In contrast, the theoretical investigations of metal-alkali molecules for the eEDM searches are limited to the work of Meyer *et al.* [22], and our recent work on Hg-alkalis (HgA) [23]. In the former, potential energy surfaces (PES) and molecular properties of Yb-alkali and Yb-alkali-earth-metal molecules are calculated at the non-relativistic level [22]. The latter involves calculations of $E_{\text{eff}}$, $W_s$ and the molecular permanent electric dipole moment (PDM) for HgA systems using Dyall cv3z basis set using a relativistic coupled cluster singles and doubles (CCSD) approach [23]. The work also presents a preliminary estimate of the expected sensitivity in eEDM experiments using HgA molecules.

Metal-alkali systems have van der Waals-like bonding, which is different from other candidate molecules with ionic bonding (e.g., ThO [6,8], HfF$^+$ [7], and YbF [24,25]). Hence, surveys for the basis set dependence and the mechanism of enhancement for $E_{\text{eff}}$ and $W_s$ for metal-alkali molecules are important, both from the viewpoint of an accurate determination of these factors, as well as for the search for good candidate molecules for eEDM searches.

In this paper, we focus on the analytical and methodological aspects of the calculations of $E_{\text{eff}}$ and $W_s$ of HgA and SrA (A = Li, Na, and K) molecules, which could be relevant for experiments with the aforementioned ultracold molecules. We summarize the three topics that we discuss in this work: (i) We calculated the $E_{\text{eff}}$, $W_s$ and the PDM of HgA (A = Li, Na, and K) molecules at the DF (Dirac-Fock) and CCSD levels of theory, using a series of basis sets from Dyall's database. We then compare our results and assess the basis sets that would be suitable for proposing eEDM candidates in these class of molecules; (ii) We study the mechanism for enhancement of $E_{\text{eff}}$ and $W_s$ in metal-alkali molecules (HgA and SrA), and compare them with that in metal-fluorides (HgF and SrF) and metal-hydrides (HgH and SrH). Our results show that although HgA has much smaller values of $E_{\text{eff}}$ and $W_s$ compared with HgH and HgF, the values of $E_{\text{eff}}$ and $W_s$ for SrLi are comparable with SrH and SrF. We explain our results by invoking the orbital interaction theory, as explained in our previous work [26]; (iii) We observe that the ratio between $W_s$ and $E_{\text{eff}}$ ($W_s/E_{\text{eff}}$) of HgX and SrX (X = H, Li, F, Na, and K) are almost constant and are independent of X. The ratio $W_s/E_{\text{eff}}$ of HgX and SrX are not significantly affected by correlation effects, which was also observed in our previous work [27]. We explain the reason for this trend by expanding $W_s/E_{\text{eff}}$ using a second quantized formalism.

## II. Theory

The expression for the eEDM operator is given by [28]

$$\hat{H}_{\text{eEDM}} = -d_e \sum_j^{N_e} \beta \mathbf{\Sigma}_j \cdot \mathbf{E}_{\text{int}}^j. \quad (1)$$

Here, $d_e$ is the eEDM, $j$ is the summation index over electronic coordinates, $N_e$ is the number of electrons in the molecule, $\beta$ is the Dirac matrix, and $\mathbf{\Sigma}$ is the four-component Pauli matrix. $\mathbf{E}_{\text{int}}$ is the internal electric field in the molecule. The effective electric field ($E_{\text{eff}}$) is given by

$$E_{\text{eff}} = -\left\langle \Psi \left| \frac{\hat{H}_{\text{eEDM}}}{d_e} \right| \Psi \right\rangle, \quad (2)$$

where $\Psi$ is the four-component electronic wavefunction of the molecule. In this work, we employed a summation over the one-electron operator for the expectation value, as given below [29,30]

$$E_{\text{eff}} = -2ic \left\langle \Psi \left| \sum_j^{N_e} \beta \gamma_5 \mathbf{p}_j^2 \right| \Psi \right\rangle, \quad (3)$$

where $i$ is the imaginary unit, $c$ is the speed of light, $\gamma_5$ is the product of Dirac matrices, and $\mathbf{p}$ is the momentum operator. The expectation value of Eq. (3) is equal to that of Eq. (1) only when $\Psi$ is the exact eigenfunction of the unperturbed Hamiltonian, which is the Dirac-Coulomb (DC) Hamiltonian, in this work.

The S-PS interaction is defined by the following operator [9,10]

$$\hat{H}_{\text{S-PS}} = \sum_A^{N_n} \hat{H}_{\text{S-PS},A}$$
$$= \sum_A^{N_n} i \frac{G_F}{\sqrt{2}} k_{s,A} Z_A \sum_j^{N_e} \beta \gamma_5 \rho_A(\mathbf{r}_{Aj}) \quad (4)$$

where $G_F$ is the Fermi coupling constant, expressed in atomic units ($2.22249 \times 10^{-14} E_h \cdot a_0^3$). $N_n$ represents the total number of the nuclei in the molecule, and $A$ labels the nuclei. $Z$ is the nuclear charge. $k_{s,A}$ is the dimensionless S-PS interaction constant of the atom $A$. We used the same Gaussian-type distribution function, for the nuclear charge density $\rho$, as in our previous work [31]. The S-PS coefficient $W_{s,A}$ is defined for molecules with $^2\Sigma$ character as follows:

$$W_{s,A} = 2\left\langle \Psi \left| \frac{\hat{H}_{\text{S-PS},A}}{k_{s,A}} \right| \Psi \right\rangle. \tag{5}$$

The permanent molecular electric dipole moment (PDM) is defined by

$$\text{PDM} = -\left\langle \Psi \left| \sum_j^{N_e} \mathbf{r}_j \right| \Psi \right\rangle + \sum_A^{N_n} Z_A \mathbf{R}_A. \tag{6}$$

Here, **r** and **R** are the position vectors of the electrons and nuclei, respectively.

We employed a relativistic CCSD method [32,33] using the DF wavefunction as the reference state. For the calculation of the expectation value of $\hat{O}$ at the CCSD level, we incorporate only the linear terms in the CCSD wave function as given below [34]

$$\left\langle \psi_0 \left| (1+\hat{T}_1+\hat{T}_2)^\dagger \hat{O}_N (1+\hat{T}_1+\hat{T}_2) \right| \psi_0 \right\rangle_C + O_0, \tag{7}$$

where $\hat{O}_N$ is the normal-ordered version of the operator, the subscript $C$ refers to connected terms, and $O_0$ is the expectation value of the operator $\hat{O}$ at the DF level [35,36].

### III. Computational Method

We use the UTChem program [37] for generating the Dirac-Fock orbitals and the molecular orbital integral transformation [38]. We use the DIRAC08 program [39] for obtaining the CCSD wave function. We modified the above-mentioned codes to calculate $E_{\text{eff}}$ [30] and $W_s$ [31]. We employ Dyall 2z, 3z, v3z, cv3z, and 4z basis sets [40–43] with polarization functions added to them in the uncontracted form for all of the elements in our target molecules. Here, v3z and cv3z refer to the basis sets of the same name, as shown in the basis set repository in DIRAC code. Table I summarizes the basis sets used in this work. Here, Dyall 2z, 3z and 4z refers to basis sets without the polarization functions [44], while Dyall 2z_pol and 4z_pol means that we added polarization functions of Dyall v3z basis sets ($g$ exponents for Hg, $d$ and $f$ exponents for Li, Na, and K, respectively) to Dyall 2z and 4z basis sets, respectively. Comparing the results obtained by using these basis sets, we shall discuss the basis set dependence of the molecular properties in HgA. All the electrons in the molecules were excited, while the virtual orbitals at higher energies were cut-off at the integral transformation and the CCSD level. The threshold energies for the cut-off are summarized in the supplemental material [45].

We use the following bond lengths (in Å); HgH: 1.7662 [46], HgF: 2.00686 [47], HgLi: 2.92 [48], HgNa: 3.52 [48] HgK: 3.90 [48], SrH: 2.1456 [46], SrF: 2.07537 [46], SrLi: 3.545 [49], SrNa: 3.889 [50], and SrK: 4.528 [50], respectively. For $W_s$, we provide only the contributions of Sr and Hg atoms, because the contribution of the lighter element for both the molecules is insignificant. We choose the following isotopes, $^{202}$Hg, and $^{88}$Sr; and employ the experimental root-mean-square charge radii [51].

## IV. Results

Hereafter, we only present the absolute values of $E_{\text{eff}}$ and $W_s$ for simplicity, while the values of PDM are shown with their sign.

### A. Basis Set Dependence

Table II shows the results for HgLi, HgNa, and HgK, at the DF level. From the table, we observe that the dependence of $E_{\text{eff}}$, $W_s$, and PDM on basis sets is very weak at the DF level. We plot the values of $E_{\text{eff}}$ and $W_s$ versus basis for HgLi, HgNa, and HgK at the CCSD level in Fig. 1. Their values are shown in Tables S1-S3 of the supplemental material [45]. From Fig. 1, we observe three common features in $E_{\text{eff}}$ and $W_s$ of HgA molecules: (i) the values obtained using the 4z basis set are not close to those obtained from the 4z_pol ones. This indicates that polarization functions play an important role in $E_{\text{eff}}$ and $W_s$ of HgA molecules. We can also see the contribution of the polarization functions from the large difference between the values at the 2z and 2z_pol; (ii) the values at the v3z, cv3z, and 4z_pol levels broadly agree with each other. From the comparison between the values at v3z and cv3z, the values of $E_{\text{eff}}$ and $W_s$ approach convergence at the cv3z level. From the small difference between v3z and 4z_pol results, we conclude that the use of 3z basis set for the occupied orbitals ($s$, $p$, $d$, $f$ for Hg, and $s$, $p$ for alkali) would be reasonably fine. In our previous work [23], we reported the error in $E_{\text{eff}}$ and $W_s$ due to the basis set at about 15% (using Dyall cv3z basis sets). However, from these figures, the error in the results from cv3z would be much smaller than 15%; (iii) the values at the 2z_pol are clearly far from those at the v3z, cv3z, and 4z_pol. From this, we understand that the 2z basis sets are not sufficient for an accurate calculation, even if we include polarization functions. However, the trends in $E_{\text{eff}}$ and $W_s$ of HgA systems are the same at any levels of basis (HgLi > HgNa > HgK). This indicates that even the 2z basis sets would be sufficient, for a qualitative analysis of $E_{\text{eff}}$ and $W_s$.

Finally, we note that the above points (ii) and (iii) are consistent with the work of Hao *et al*. [52]. They calculated the $P$-odd interaction coefficient $W_A$ for BaF by employing the relativistic coupled cluster method. In their work, the values they obtained using Dyall v2z are clearly far from those with Dyall v3z basis, while the difference between the values at the Dyall v3z and Dyall v4z is not too significant. The similarities between their results for BaF and our results for HgA indicates that the dependence of these properties using Dyall basis sets would not significantly depend on the electronic structure of molecules.

In Fig. 2, we plot the values of PDM for HgA at the CCSD level, whose numerical values are shown in tables S1-S3 in the supplemental material [45]. The direction of the PDM is taken along the molecular axis from the mercury to the alkali atom. The basis set dependence of PDM is similar qualitatively to that observed in $E_{\text{eff}}$ and $W_s$. Also, the basis set dependence of PDM is stronger than that observed in $E_{\text{eff}}$ and $W_s$; e.g., values at the 2z, 3z, and 2z_pol basis sets do not reproduce the sign of 4z_pol. In contrast to this strong dependence, the values at the v3z, cv3z, and 4z_pol are in broad agreement, similar to $E_{\text{eff}}$ and $W_s$ in Fig. 1. We, therefore, assess that the results are extremely sensitive to basis sets only for low-quality basis sets (e.g., 2z quality, with and without polarization functions), and hence our previous calculation of PDM at the cv3z [23] are sufficiently accurate, at least from the point of view of proposing new candidates for eEDM search experiments.

### B. Analysis of $E_{\text{eff}}$ and $W_s$

In this section, we discuss why HgA has much smaller values of $E_{\text{eff}}$ and $W_s$ than HgH [26,53] and HgF [26,54] based on the electronic structure of these molecules. The dependence of molecular enhancement factors of $P$ and $P$, $T$-odd violating properties on the nuclear charge have been investigated thoroughly (e.g., Refs. [52,55–57]). However, the small values of HgA cannot be explained only by invoking nuclear charge. We show the results of the Mulliken population (MP) analysis [58] for the singly occupied molecular orbital (SOMO) in Table III. We employ the Dyall cv3z basis sets for the MP calculations (note that the values of MP for HgH and HgF are not exactly same as our previously reported values [26], where 4z quality basis set was employed). The value of MP indicates the contribution of each atomic orbital to SOMO in the target molecule. From Table III, we see that the SOMO electrons are localized in the Hg atom for both HgH and HgF. In contrast, the SOMO for HgA is not localized in Hg, but the alkali atom. Since the SOMO is not localized to Hg, HgA molecules do not have an enhanced $E_{\text{eff}}$ and $W_s$ that would have resulted from the large nuclear charge of Hg ($Z = 80$). This shows the reason why HgA has much smaller $E_{\text{eff}}$ and $W_s$ than HgH and HgF, although they contain the Hg atom.

Although the HgA molecules have much smaller $E_{\text{eff}}$ and $W_s$ than HgH and HgF, metal-alkali molecules need not always have a small $E_{\text{eff}}$ and $W_s$. We present the results of SrH, SrF, and SrA (A = Li, Na, and K) in Table IV. We employ the Dyall cv3z basis sets for all the elements. The virtual orbitals are cut-off at 100 a.u. The trends in $E_{\text{eff}}$ and $W_s$ for Sr molecules is SrH > SrF > SrLi > SrNa > SrK, which are same as those found in Hg-containing molecules. However, the values of $E_{\text{eff}}$ and $W_s$ for SrLi are almost the same as those of SrF, at both the DF and the CCSD levels. SrNa and SrK have $E_{\text{eff}}$ and $W_s$ that are about one order smaller at the DF level, but the values at the CCSD level are about a half and a third of SrH, respectively. It is in stark contrast to HgNa and HgK; $E_{\text{eff}}$ and $W_s$ of HgNa and HgK are one-sixth and one-seventh smaller than HgH at the CCSD level, respectively. Although $E_{\text{eff}}$ and $W_s$ of SrA are smaller than SrH and SrF, the decrease in $E_{\text{eff}}$ and $W_s$ for SrA is clearly lesser than that in HgA. As a result, SrA has enough large $E_{\text{eff}}$ and $W_s$ to be proposed for an eEDM experiment.

The reason why SrA possess relatively large $E_{\text{eff}}$ and $W_s$ could be explained by invoking the orbital interaction theory [59–61]. Fig. 3 shows the energy diagram for the atomic and molecular orbitals of HgLi, HgK, SrLi, and SrK. Here, the atomic orbital energies are obtained from atomic DF calculations using the GRASP2K code [62].

We omit HgNa (SrNa), because their electronic structures are in between HgLi (SrLi) and HgK (SrK).

From Fig. 3, we observe that the 6$s$ electron of Hg hardly transfers to the alkali, because Hg's 6$s$ orbital is more stable than the valence $s$ orbitals of alkalis, and the transfer would lead to an energetic instability (note that if the electron transfer did not occur at all, $E_{\text{eff}}$ and $W_s$ of HgA would become zero, because Hg is a closed shell system). As a result, in HgLi and HgK, the 2$s$ and 4$s$ orbitals of Li and K mainly contribute to the SOMO, while 6$s$ of Hg mainly contributes to SOMO-1. These electronic structures are in contrast with HgH and HgF molecules, where Hg's 6$s$ and 6$p$ mainly contributes to SOMO (see Fig. 1 in Ref. [26]).

In contrast, the valence 5$s$ of Sr is unstable as compared to 6$s$ of Hg. The valence atomic orbitals in Sr and alkali mix more than in the case of Hg and alkali; i.e., the chemical bonds of SrA would not be completely van der Waals like, and it is more covalent than HgA. The difference between the valence $s$ orbitals of Hg and Sr can be explained on account of the stabilization of Hg's 6$s$, which is due to the relativistic contraction effect and the weak screening effect of 5$d$ electrons. As a result, the 5$s$ of Sr can contribute to SOMO more than the 6$s$ of Hg in HgA. Especially, in the case of SrLi, the 2$s$ of Li is slightly more stable than 5$s$ of Sr, and the electronic structure of SrLi is similar to that of HgH (Fig. 1 in Ref. [26]), rather than that of HgA. Therefore, $E_{\text{eff}}$ and $W_s$ of SrLi are similar to those of SrH and SrF. As an aside, we note that HgA has larger $E_{\text{eff}}$ and $W_s$ than SrA due to the larger $Z$ and relativistic effect of Hg, despite the small contribution of Hg to SOMO.

We show the values of the differences in the energies of atomic orbitals (AO), as well as the overlap integrals between the 5s and 6s orbitals of Sr and Hg atoms and the valence orbitals of the alkalis for HgA and SrA in Table V. The overlap integrals were obtained by using the contracted Dyall 4z basis sets. We have already discussed the energy differences between the AOs in the previous paragraphs using Fig. 3. The values of the overlap integrals of HgA are clearly smaller when compared to SrA with same alkalis. It is consistent with the discussion presented above that the chemical bonds of SrA can become more covalent than those of HgA, due to the contraction of the 6$s$ orbital of Hg.

### C. Ratio $W_s/E_{\text{eff}}$

The ratios ($W_s/E_{\text{eff}}$) for atoms were first estimated by Dzuba *et al.* [63,64]. Gaul *et al.* studied molecular $W_s/E_{\text{eff}}$ systematically [57], and mentioned that the ratio is rather insensitive to the "chemical environment" of the heavy nucleus. However, it is unclear if their conclusion can be extended to HgA whose electronic structures are significantly different from Gaul *et al.*'s target molecules; hydrides, nitrides, oxides, and fluorides.

Table IV shows the values of $W_s/E_{\text{eff}}$ for HgX and SrX (X = H, Li, F, Na, and K) at the DF and CCSD levels. From this table, we observe that $W_s/E_{\text{eff}}$ are almost same at both DF and CCSD levels for HgX and SrX, although the $E_{\text{eff}}$ and $W_s$ are different between the alkalides, the hydrides, and the fluorides.

We can understand the reason for the weaker dependence of the ratio on the molecular electronic structure, by using analytical formulae for $E_{eff}$ and $W_s$. First, we shall discuss this aspect only at the DF level. Here, we use the expression for $E_{eff}$ and $W_s$ based on the molecular orbital representation proposed by Meyer *et al*. According to their approximation, $E_{eff}$ is expressed as follows in atomic units [65,66]

$$E_{eff} = \frac{4\sigma}{\sqrt{3}} c_s c_p Z \Gamma_{rel}, \quad (8)$$

where

$$\Gamma_{rel} = -\frac{4Z^2 \alpha^2 Z_i^2}{\gamma(4\gamma^2 - 1)(\nu_s \nu_p)^{3/2}}, \quad (9)$$

and

$$\gamma = \sqrt{(j+1/2)^2 - (Z\alpha)^2}. \quad (10)$$

Here, $\sigma = \pm 1/2$, which is related to the projection of the spin to the molecular axis. $\alpha$ is the fine structure constant. $Z_i$ is the effective nuclear charge seen by the valence electron; for a neutral atom $Z_i = 1$. $\nu$ is the effective quantum number, and $j$ is the total angular momentum. $c_s$ and $c_p$ are the molecular orbital coefficients in SOMO ($\psi_{SOMO}$) and are represented as follows

$$|\psi_{SOMO}\rangle = c_s |s\rangle + c_p |p\rangle + \sum_{other} c_{other} |other\rangle, \quad (11)$$

where

$$|p\rangle = -\frac{2\sigma}{\sqrt{3}} |p_{1/2}\rangle + \sqrt{\frac{2}{3}} |p_{3/2}\rangle. \quad (12)$$

Here, $|s\rangle$ and $|p\rangle$ is the valence $s$ and $p$ orbitals of the heavier atom in the molecule, which mainly contributes to $E_{eff}$. $|other\rangle$ refers to atomic orbitals excluding $|s\rangle$ and $|p\rangle$.

Eq. (8) can be rewritten such that the contributions from the nuclear charge $Z$ ($\Gamma'_{rel,edm}$) and the electronic structure of the molecule ($X$) are separated, as given below

$$E_{eff} = X\Gamma'_{rel,edm}, \quad (13)$$

$$X = \frac{4\sigma}{\sqrt{3}} \frac{c_s c_p}{(\nu_s \nu_p)^{3/2}}, \quad (14)$$

and

$$\Gamma'_{rel,edm} = -\frac{4Z^3 \alpha^2 Z_i^2}{\gamma(4\gamma^2 - 1)}. \quad (15)$$

Next, we give the analogous expression to $E_{eff}$ in the case of $W_s$ for $^{1/2}\Sigma$ molecule as follows, using the following atomic expression of $W_s$ [2,9]

$$W_s = 2X\Gamma'_{rel,S-PS}, \quad (16)$$

and

$$\Gamma'_{rel,s-ps} = \frac{G_F Z^3 \alpha}{\pi\sqrt{2}} \frac{4\gamma^2}{\Gamma^2(2\gamma+1)} \left(\frac{1}{2ZR_{nuc}}\right)^{2-2\gamma}. \quad (17)$$

Here, $R_{nuc}$ is the nuclear radius. We modify Eq. (8) in Ref. [9] so that it is consistent with the expression for $X$ given in Eq. (14) for $W_s$. Eqs. (13) and (16) are based on the first-order perturbation theory and are represented as the expectation values of the unperturbed wavefunction. It is a good approximation because the influence of the eEDM and the S-PS interactions on the wavefunction should be negligible compared with that of the unperturbed

Hamiltonian.

To get insights into the ratio $W_s/E_{\text{eff}}$, it is important that the part depending on the electronic structure of the molecules ($X$ in Eq. (14)) is common for both $E_{\text{eff}}$ and $W_s$ and hence cancel each other out in $W_s/E_{\text{eff}}$. As a result, the remaining part of $W_s/E_{\text{eff}}$ depends on the only $Z$, as follows

$$W_s/E_{\text{eff}} \propto \frac{\gamma^2}{\Gamma^2(2\gamma+1)} \left(\frac{1}{2ZR_{\text{nuc}}}\right)^{2-2\gamma} \Big/ \frac{1}{\gamma(4\gamma^2-1)}. \tag{18}$$

Here, we ignore some of the coefficients and physical constants that are not relevant for the analysis, for simplicity. From the expression given above, we can explain the reason why each of the HgX and SrX has similar $W_s/E_{\text{eff}}$ at the DF level and is found to be due to cancelation between the parts corresponding to the molecular electronic structure.

From Eq. (18), $W_s/E_{\text{eff}}$ increases as $Z$, because $(1/[2ZR_{\text{nuc}}])^{2-2\gamma}$ in the numerator increases faster than the denominator. It is consistent with the fact that relativistic effects in the S-PS interaction are larger than that in the eEDM. The former is the interaction between nucleons and the electrons inside the nucleus, while the latter is between the nuclear charge and the electrons distributed close to the nucleus.

Another observation from Table IV is that the values of $W_s/E_{\text{eff}}$ at the DF and CCSD levels are almost the same, although each value of $E_{\text{eff}}$ and $W_s$ is different due to correlation effects.

We explain this trend by utilizing the representation of the one-electron operator in the second quantized form. A one-electron operator $\hat{O}$ can be given by [67],

$$\hat{O} = \sum_{q,r} O_{qr} a_q^\dagger a_r, \tag{19}$$

where

$$O_{qr} = \int \psi_q^* \hat{O} \psi_r d\tau. \tag{20}$$

In the above expression, $\psi$ refers to the one-electron molecular spin-orbitals, $q$ and $r$ are arbitrary indices of the spin-orbitals, $a_q^\dagger$ ($a_r$) is an electron creation operator (annihilate operator) in a spin-orbital $\psi_q$ ($\psi_r$), respectively. From Eq. (19), the expectation value of $E_{\text{eff}}$ and $W_s$ can be written as follows

$$\begin{aligned} E_{\text{eff}} &= \langle \Psi | \hat{E}_{\text{eff}} | \Psi \rangle \\ &= \sum_{q,r} (E_{\text{eff}})_{qr} D_{qr}, \end{aligned} \tag{21}$$

$$\begin{aligned} W_s &= \langle \Psi | \hat{W}_s | \Psi \rangle \\ &= \sum_{q,r} (W_s)_{qr} D_{qr}, \end{aligned} \tag{22}$$

where

$$(E_{\text{eff}})_{qr} = \int \psi_q^* \hat{E}_{\text{eff}} \psi_r d\tau, \tag{23}$$

$$(W_s)_{qr} = \int \psi_q^* \hat{W}_s \psi_r d\tau, \tag{24}$$

and

$$D_{qr} = \langle \Psi | a_q^\dagger a_r | \Psi \rangle. \tag{25}$$

Here, $\Psi$ is the electronic wavefunction of the unperturbed Hamiltonian, which includes the correlation effects.

Next, we generalize Eq. (11) for specifically highlighting the inner $s$ and $p$ orbitals. First, $\psi_q$

may be expressed as follows

$$|\psi_q\rangle = \sum_{s'} c_{s'q}|s'\rangle + \sum_{p'} c_{p'q}|p'\rangle + \sum_{\text{other}} c_{\text{other}}|\text{other}\rangle. \tag{26}$$

Here, $s'$ and $p'$ are the indices of the heavier atomic $s$ and $p$ orbitals in the molecule, respectively. Replacing Eq. (11) with Eq. (26), we can express Eq. (8) as follows

$$(E_{\text{eff}})_{qr} \approx X_{qr}\Gamma'_{\text{rel,edm}}, \tag{27}$$

and

$$X_{qr} = \frac{2\sigma}{\sqrt{3}} \sum_{s',p'} \frac{1}{(\nu_{s'}\nu_{p'})^{3/2}} \left( c^*_{s'q}c_{p'r} + c^*_{p'q}c_{s'r} \right). \tag{28}$$

From Eq. (21), $E_{\text{eff}}$ is expressed as follows

$$E_{\text{eff}} \approx \Gamma'_{\text{rel,edm}} \sum_{q,r} X_{qr} \langle \Psi | a_q^\dagger a_r | \Psi \rangle. \tag{29}$$

The above derivation using Eq. (26) can also be extended to $W_s$, and one obtains

$$W_s \approx \Gamma'_{\text{rel,s-ps}} \sum_{q,r} X_{qr} D_{qr}. \tag{30}$$

Finally, $W_s/E_{\text{eff}}$ can be expressed as follows

$$W_s/E_{\text{eff}} \approx \frac{\Gamma'_{\text{rel,s-ps}} \sum_{q,r} X_{qr} D_{qr}}{\Gamma'_{\text{rel,edm}} \sum_{q,r} X_{qr} D_{qr}} = \frac{\Gamma'_{\text{rel,s-ps}}}{\Gamma'_{\text{rel,edm}}}$$

$$\propto \frac{\gamma^2}{\Gamma^2(2\gamma+1)} \left( \frac{1}{2ZR_{\text{nuc}}} \right)^{2-2\gamma} \Big/ \frac{1}{\gamma(4\gamma^2-1)} \tag{31}$$

We see that the parts that depend on the electronic structure $\left( \sum_{q,r} X_{qr} D_{qr} \right)$ cancel out. As a result, $W_s/E_{\text{eff}}$ depends on only the nuclear charge $Z$ even at the correlation level.

The points mentioned above can explain the trend in our previously reported results [27]; $E_{\text{eff}}$ and $W_s$ in HgF are each larger than that of RaF (i.e., Z independent), while the ratio ($W_s/E_{\text{eff}}$) is larger for RaF (i.e., monotonically $Z$ dependent). The values of $E_{\text{eff}}$ and $W_s$ themselves depend on the contraction of the core region of the outermost orbitals [27], and reflect the electronic structure of molecular orbitals. In contrast, $W_s/E_{\text{eff}}$ only depends on $Z$ and $R_{\text{nuc}}$ as shown in Eqs. (18) and (31), and are independent of the electronic structure of the molecule.

In the above discussion, we ignore the effects of $p_{3/2}$, $d$, $f$, ... orbitals. However, their contributions to $E_{\text{eff}}$ and $W_s$ are much smaller than those from the $s$ and $p_{1/2}$ orbitals, because these orbitals are not distributed significantly in the region close to the nucleus. Hence, even though the excitations from (to) $p_{3/2}$, $d$, $f$, ... orbitals may exist at the correlation levels, their effects on the total value of $W_s/E_{\text{eff}}$ would be insignificant, as our results for the ratio at the CCSD level indicate. Our analysis can be generalized to any molecules whose $s$ and $p_{1/2}$ orbitals mainly contribute to their open-shell configurations. For example, it is reported that the ratio $W_s/E_{\text{eff}}$ of triatomic molecules are almost the same as that of the corresponding monofluorides [68]. This result can also be understood from our analysis.

## VI. CONCLUSION

To conclude, we have discussed three broad aspects of HgA as well as other systems that could be of interest for future eEDM search experiments. In the first part of this work, we calculate molecular properties of HgA, using a series of basis sets from Dyall's database. We find that double zeta basis sets do not lead to the correct sign for the PDM, even if polarization functions are included. In contrast, the cv3z basis set is of reasonably good quality for the calculation of $E_{\text{eff}}$, $W_s$, and PDM, as are the v3z and 4z_pol ones. This survey will help in the correct choice of basis for calculations that are of interest to fundamental physics, involving molecules with van der Waals-like character, such as HgA.

Next, we analyze the reason for smaller values of $E_{\text{eff}}$ and $W_s$ in HgA systems, where due to the Hg atom, we would normally expect a large $E_{\text{eff}}$ and $W_s$ (for example, HgX and HgH). We found that in these systems, where van der Waals-like character is present, the SOMO electron localizes in the alkali atom, and leads much smaller $E_{\text{eff}}$ and $W_s$. We also observed that the values of $E_{\text{eff}}$ and $W_s$ for SrLi are comparable with SrH and SrF. The difference between HgA and SrA can be attributed to the stabilization of the valence 6s orbital of Hg. Our idea that SrA has relatively large $P$, $T$-odd properties could be extended to molecules containing the same group-2 systems; e.g., BaA and RaA.

Lastly, we found that the ratio $W_s/E_{\text{eff}}$ hardly depends on the electronic structure and the correlation effects, and dominantly depends on $Z$. We explain the reason for this behavior is due to the cancellation of the electronic structure parts in $W_s/E_{\text{eff}}$. This analysis supports the suggestion of our [27] and Gaul *et al.*'s [57] previous works that performing two different experiments using heavy molecules, and relatively light molecules, is important to separate the contribution from the eEDM and the S-PS interactions. Based on these points, and in view of the successes in ultracold molecules that have been already reported using Sr containing systems (e.g., SrF [69]), in combination with the suitability of SrA for laser cooling as discussed in Ref. [70], we propose that the combination of SrA and one of the current leading candidate molecules that are heavier (ThO and HfF[+]) to separate the eEDM and the S-PS interactions.

## V. ACKNOWLEDGMENTS

We would like to thank Profs. B. P. Das, A. C. Vutha, and K. Dyall for valuable discussions. This work was supported by JSPS KAKENHI Grant No. 17J02767, No. 17H03011, No. 17H02881, and No. 18K05040.

Table I Basis set information.

|        | Hg               | Li      | Na         | K          | Sr[b]           | H[b]    | F[b]       |
|--------|------------------|---------|------------|------------|-----------------|---------|------------|
| 2z     | 24s19p12d8f      | 10s6p   | 12s8p      | 15s11p     | -               | -       | -          |
| 3z     | 30s24p15d10f     | 14s8p   | 18s11p     | 23s16p     | -               | -       | -          |
| 4z     | 34s30p19d12f     | 18s10p  | 24s14p     | 30s21p     | -               | -       | -          |
| 2z_pol | 24s19p12d8f2g    | 10s6p2d | 12s8p4d2f  | 15s11p4d2f | -               | -       | -          |
| v3z    | 30s24p15d11f2g   | 14s8p2d | 19s12p4d2f | 24s17p4d2f | -               | -       | -          |
| cv3z   | 30s24p15d11f4g1h | 14s8p2d | 19s12p5d2f | 24s17p6d3f | 29s20p13d5f1g   | 9s2p1d  | 14s8p3d1f  |
| 4z_pol | 34s30p19d12f2g   | 18s10p2d| 25s14p4d2f | -[a]       | -               | -       | -          |

[a] We did not employ 4z_pol basis set for HgK.
[b] We employed only cv3z basis set for Sr, H, and F atoms.

Table II Results of our calculations at the DF level.

|           | $E_{\text{eff}}$ (GV/cm) | | | $W_s$ (kHz) | | | PDM(D) | | |
|-----------|------|------|------|------|------|------|-------|-------|-------|
| Basis set | HgLi | HgNa | HgK  | HgLi | HgNa | HgK  | HgLi  | HgNa  | HgK   |
| 2z        | 13.39| 7.45 | 5.70 | 29.51| 16.41| 12.56| -1.58 | -0.94 | -1.59 |
| 3z        | 13.76| 7.61 | 5.82 | 31.08| 17.19| 13.15| -1.45 | -0.85 | -1.42 |
| 4z        | 13.77| 7.62 | 5.82 | 32.29| 17.23| 13.17| -1.44 | -0.84 | -1.39 |
| 2z_pol    | 13.44| 7.46 | 5.61 | 31.13| 16.44| 12.34| -1.58 | -0.96 | -1.60 |
| v3z       | 13.74| 7.59 | 5.73 | 29.63| 17.15| 12.95| -1.47 | -0.88 | -1.48 |
| cv3z      | 13.74| 7.59 | 5.73 | 31.02| 17.15| 12.95| -1.47 | -0.88 | -1.48 |
| 4z_pol    | 13.75| 7.60 | -    | 31.02| 17.18| -    | -1.45 | -0.86 | -     |

Table III Mulliken population analysis for HgX (X = H, Li, F, Na and K) using Dyall cv3z basis sets.

|          | HgH              | HgF              | HgLi  | HgNa  | HgK   |
|----------|------------------|------------------|-------|-------|-------|
| Hg ($s$) | 0.42             | 0.74             | -0.01 | -0.01 | -0.01 |
| Hg ($p$) | 0.38             | 0.16             | 0.10  | 0.06  | 0.06  |
| Hg total | 0.82             | 0.94             | 0.10  | 0.05  | 0.04  |
| X ($s$)  | 0.18             | $-8 \times 10^{-4}$ | 0.87  | 0.94  | 0.94  |
| X ($p$)  | $-1 \times 10^{-3}$ | 0.06          | 0.03  | 0.01  | 0.02  |
| X total  | 0.18             | 0.06             | 0.90  | 0.95  | 0.96  |

Table IV Results of our calculations for $E_{eff}$ (GV/cm), $W_s$ (kHz) and $W_s/E_{eff}$ (kHz/(GV/cm)) using Dyall cv3z basis set.

| Method | Property | SrH | SrF | SrLi | SrNa | SrK | HgH | HgF | HgLi | HgNa | HgK |
|---|---|---|---|---|---|---|---|---|---|---|---|
| DF | $E_{eff}$ | 1.76 | 1.50 | 1.41 | 0.41 | 0.24 | 106.79 | 105.32 | 13.74 | 7.59 | 5.73 |
| DF | $W_s$ | 1.59 | 1.35 | 1.27 | 0.37 | 0.22 | 241.72 | 237.71 | 31.02 | 17.15 | 12.95 |
| DF | $W_s/E_{eff}$ | 0.90 | 0.90 | 0.90 | 0.89 | 0.92 | 2.26 | 2.26 | 2.26 | 2.26 | 2.26 |
| CCSD | $E_{eff}$ | 2.61 | 2.10 | 2.02 | 1.48 | 0.97 | 120.13 | 115.73 | 37.79 | 20.33 | 16.24 |
| CCSD | $W_s$ | 2.33 | 1.88 | 1.81 | 1.33 | 0.86 | 277.89 | 266.65 | 86.37 | 46.46 | 37.05 |
| CCSD | $W_s/E_{eff}$ | 0.89 | 0.89 | 0.90 | 0.90 | 0.89 | 2.31 | 2.30 | 2.29 | 2.29 | 2.28 |

Table V. Results for the AO energy differences and the absolute values of the overlap integral.

|  | HgLi | HgNa | HgK | SrLi | SrNa | SrK |
|---|---|---|---|---|---|---|
| Overlap integral | 0.38 | 0.29 | 0.28 | 0.42 | 0.38 | 0.34 |
| AO energy difference (a.u.) | -0.13[a] | -0.15[a] | -0.18[a] | 0.015 | $1.1 \times 10^{-3}$ | -0.03[a] |

[a] The negative sign means that the AO energy of valence $s$ orbital of Hg (Sr) is lower than alkali atom.

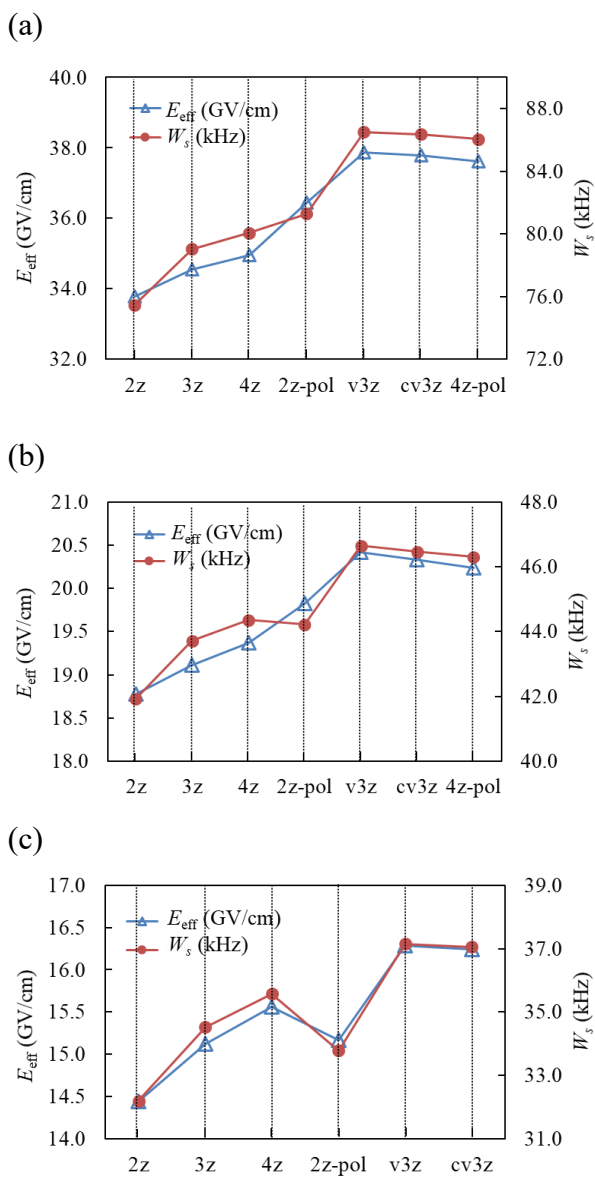

FIG. 1. Calculation results of $E_{\text{eff}}$ and $W_s$ for (a) HgLi, (b) HgNa and (c) HgK at the CCSD level.

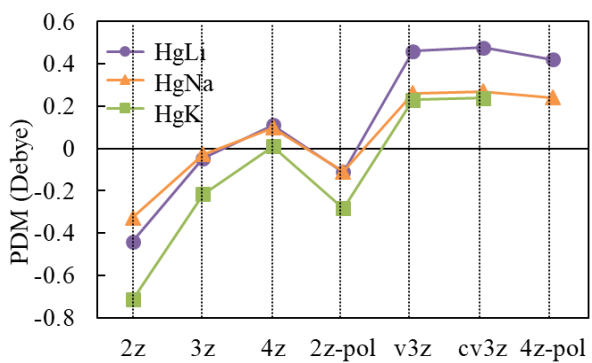

FIG. 2. Calculation results of PDM for HgLi, HgNa and HgK at the CCSD level.

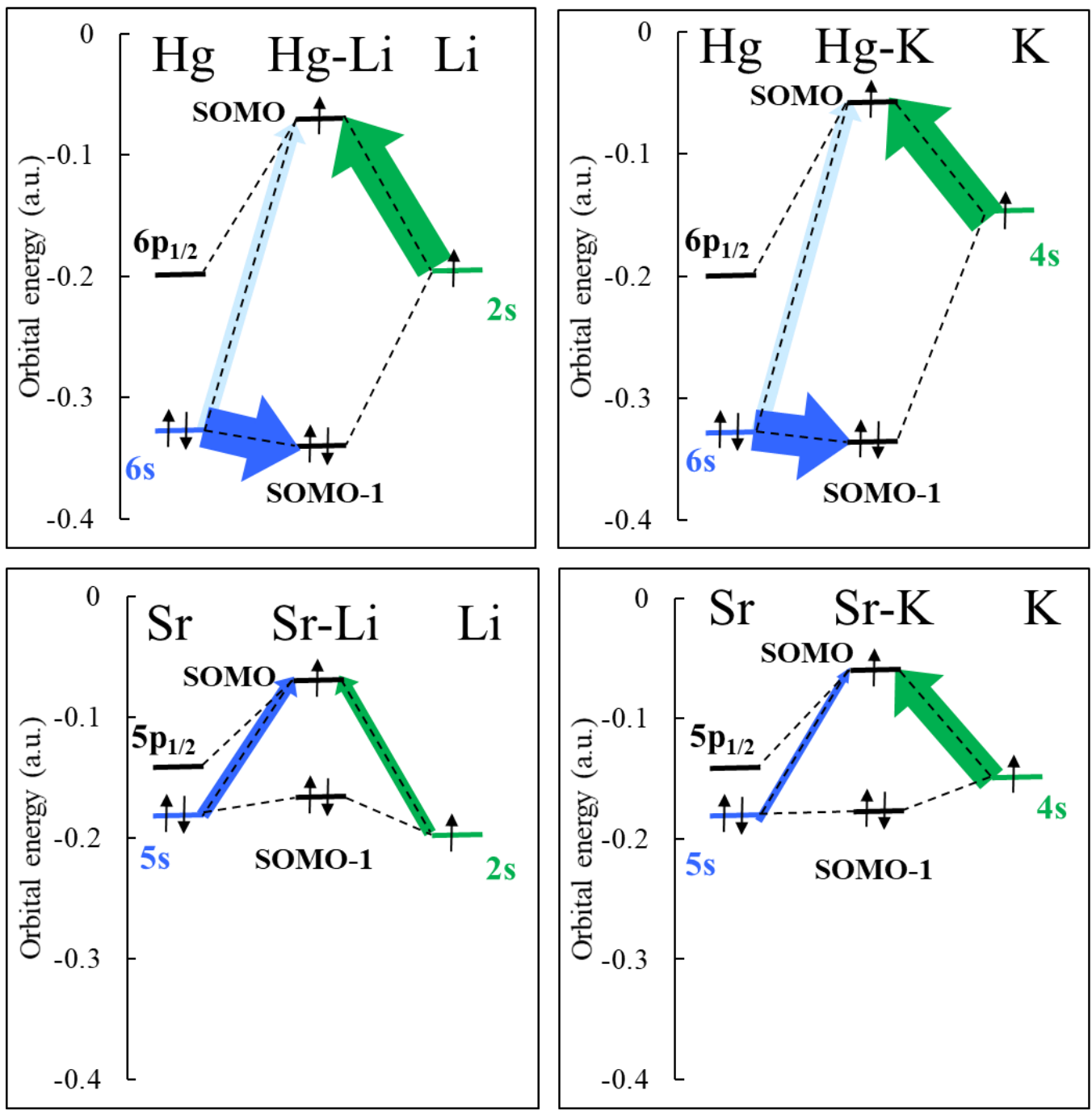

FIG. 3. Energy diagrams of HgLi, HgK, SrLi, and SrK molecules

The energies of the valence occupied orbitals of Li, K, Sr and Hg ($2s$, $4s$, $5s$, and $6s$) were evaluated from the ground states of the atoms. The $5p$ and $6p$ orbital energies of Sr and Hg were evaluated from the excited state of the atoms whose valence electron configurations are $ns^1 np^1$ ($n$ = 5, 6). The atomic calculations were based on GRASP2K [57]. MO energies of the four molecules were evaluated at the DF level and Dyall cv3z basis sets.

# Supplemental Material

Tables S1-S3 show the CCSD results for HgLi, HgNa, and HgK respectively, and also provide the information on our active space. We plot the values shown in Tables S1-S3 in Fig. 1 of the main text, and discuss these values in detail. The direction of the PDM is taken along the molecular axis from the mercury to the alkali atom.

Table S1. Results of our calculations on HgLi at the CCSD level.

| Basis set | $E_{eff}$ (GV/cm) | $W_s$ (kHz) | PDM (Debye) | virtual cutoff (a.u.) | Total Basis Spinor Sets | Active orbitals |
|---|---|---|---|---|---|---|
| 2z | 33.79 | 75.46 | -0.44 | 500 | 450 | 322 |
| 3z | 34.55 | 79.04 | -0.046 | 500 | 570 | 382 |
| 4z | 34.95 | 80.07 | 0.11 | 500 | 702 | 454 |
| 2z_pol | 36.45 | 81.28 | -0.11 | 300 | 506 | 354 |
| v3z | 37.86 | 86.52 | 0.46 | 300 | 640 | 424 |
| cv3z | 37.79 | 86.37 | 0.48 | 100 | 698 | 448 |
| 4z_pol | 37.61 | 86.05 | 0.42 | 300 | 758 | 478 |

Table S2. Results of our calculations on HgNa at the CCSD level.

| Basis set | $E_{eff}$ (GV/cm) | $W_s$ (kHz) | PDM (Debye) | virtual cutoff (a.u.) | Total Basis Spinor Sets | Active orbitals |
|---|---|---|---|---|---|---|
| 2z | 18.78 | 41.93 | -0.33 | 500 | 466 | 334 |
| 3z | 19.11 | 43.71 | -0.028 | 500 | 596 | 396 |
| 4z | 19.37 | 44.36 | 0.10 | 500 | 740 | 470 |
| 2z_pol | 19.83 | 44.22 | -0.11 | 300 | 570 | 408 |
| v3z | 20.41 | 46.65 | 0.26 | 300 | 722 | 486 |
| cv3z | 20.33 | 46.46 | 0.27 | 100 | 790 | 504 |
| 4z_pol | 20.24 | 46.30 | 0.24 | 300 | 844 | 534 |

Table S3. Results of our calculations on HgK at the CCSD level.

| Basis set | $E_{eff}$ (GV/cm) | $W_s$ (kHz) | PDM (Debye) | virtual cutoff (a.u.) | Total Basis Spinor Sets | Active orbitals |
|---|---|---|---|---|---|---|
| 2z | 14.44 | 32.19 | -0.71 | 500 | 490 | 342 |
| 3z | 15.12 | 34.52 | -0.22 | 500 | 636 | 410 |
| 4z | 15.56 | 35.57 | 0.010 | 500 | 792 | 490 |
| 2z_pol | 15.17 | 33.78 | -0.28 | 300 | 594 | 422 |
| v3z | 16.28 | 37.15 | 0.23 | 300 | 762 | 508 |
| cv3z | 16.24 | 37.05 | 0.24 | 100 | 854 | 560 |